\documentclass[12pt]{iopart}

\begin{document}

\title[Solutions of Einstein Field Equation for ...]{Solutions of Einstein Field Equation for an Extra-Dimensional Anisotropic Metric with Two Scale Factors}

\author{Taymaz Ghaneh}

\address{Research Institute for Astronomy and Astrophysics of Maragha (RIAAM)}
\ead{Ghaneh@TabrizU.ac.ir}
\vspace{10pt}
\begin{indented}
\item[]December 2016
\end{indented}

\begin{abstract}
The manuscript studies a 3+N+1-dimensional space in which the N extra dimensions are dynamically compact. The 3 large dimensions, behaving as the spacial part of the FRW metric, possess a different scale factor in comparison with the N extra ones, making the whole space anisotropic. The possible effects caused by the existence of a common time-like coordinate between the compact dimensions and our 3-dimensional hypersurface are investigated. The higher dimensional Friedmann-Like equations of the mentioned model are achieved. The continuity equation is reached at the special case of 3+4+1-dimensional metric. It is shown that not only the existence of the extra dimensions itself but also the pressure difference between the 3-dimensional hypersurface and the compact dimensions might get probed on the hypersurface as an additive source of gravity with the same behavior as baryonic matter. Furthermore, the relation between the coupling constant of the higher-dimensional universe and the Newton's constant of gravitation is investigated to reach an estimated limit for it. As another aim, the literature studies the role of dimensionality on the behavior of the higher-dimensional Friedmann equations.
\end{abstract}

%
\vspace{2pc}
%
\submitto{}
%
%
%

\section{\label{sec1}Introduction}

Extra dimensions show up in many cases in theoretical physics, such as uniting electromagnetism and gravity in Kaluza-Klein theory \cite{Kaluza,Klein} or resolving the anomaly in type I string theory by the idea of supersymmetry \cite{Green–Schwarz mechanism}. Generally speaking, it is common to categorize extra-dimensional models in to two parts, whether they contain compact extra dimensions of internal degrees of freedom or large-scale extensive dimensions of a higher-dimensional space called \emph{Hyperspace} or \emph{Bulk}. As some examples of the two cases one can mention \cite{Bringmann,Jalalzadeh,Ivashchuk,Zhuk,Huang}.\\
The concentration here is upon special prospects like dark matter and dark energy alternatives. This manuscript attempts to understand the gravitational effects which appears by considering compact extra dimensions added to the structure of our FRW universe, to be interpreted as dark matter or dark energy. These effects are researched due to two aspects: First, the influences of a shared cosmic time between our large-scale universe and the extra dimensions, and second, the role which the dimensionality of the compact part can play in these influences.\\

\section{\label{sec2}\emph{\lowercase{n}+N+1}-Dimensional Space-time}

The metric of the concerned $n+N+1-dimensional$ space-time is defined as follows, containing the scale factor $a(t)$ of the large dimensions and another scale factor $A(t)$ for the compact ones:

\begin{eqnarray}\label{metric}
&& dS^2=c^2 dt^2-a^2(t)(\frac{dr^{2}}{1-kr^{2}}+r^{2}d\Omega_{n-1}^{2})-A^{2}(t)(\frac{dR^{2}}{1-jr^2}+R^{2}d\Omega_{N-1}^{'2}) 
\end{eqnarray}

where $d\Omega_{n-1}^{2}$ is the angular part of the $n$ large spatial dimensions, and $d\Omega_{n-1}^{'2}$ is the angular part of the $N$ compact ones. $k$ and $j$ are also curvature parameters of them, respectively. Although the number of large spacial dimensions are refereed by $n$ here, $n$ is assumed equal to $3$ in the whole document, as expected by inverse-square law of gravity. In this literature, the space-like part in the metric (\ref{metric}) which possess the scale factor $a(t)$ is referred as the \emph{FRW spacial part} and the space-like part possessing the scale factor $A(t)$ is referred as the \emph{compact extra part} of the metric. Non-zero entries of the Einstein tensor of this metric in mixed basis, considering a factor $c^{2}$, are

\numparts
\begin{eqnarray}\label{Etensor}
&& E_{0}^{0}c^2=\frac{n(n-1)}{2}H_{(k)}^{2}(t)+\frac{N(N-1)}{2}H_{(j)}^{'2}(t)+nNH(t)H^{'}(t), \label{Etensor first} \\
\nonumber && \\
\nonumber && E_{1}^{1}c^2=E_{2}^{2}c^2=...=E_{n}^{n}c^2= \\
\nonumber && \frac{(n-1)(n-2)}{2}H_{(k)}^{2}(t)+\frac{N(N-1)}{2}H_{(j)}^{'2}(t)\\
&& +(n-1)NH(t)H^{'}(t)+(n-1)Q(t)+NQ^{'}(t), \label{Etensor second} \\
\nonumber && \\
\nonumber && E_{n+1}^{n+1}c^2=E_{n+2}^{n+2}c^2=...=E_{n+N}^{n+N}c^2= \\
\nonumber && \frac{n(n-1)}{2}H_{(k)}^{2}(t)+\frac{(N-1)(N-2)}{2}H_{(j)}^{'2}(t)\\
          && +n(N-1)H(t)H^{'}(t)+nQ(t)+(N-1)Q^{'}(t), \label{Etensor third}
\end{eqnarray}
\endnumparts

where

\numparts
\begin{eqnarray}\label{HQ}
&& H_{(k)}^{2}(t)=\frac{\dot{a}^{2}(t)+kc^{2}}{a^{2}(t)}, H_{(j)}^{'2}(t)=\frac{\dot{A}^{2}(t)+jc^{2}}{A^{2}(t)}\\
&& H(t)=\frac{\dot{a}(t)}{a(t)}, H^{'}(t)=\frac{\dot{A}(t)}{A(t)}, \\
          && Q(t)=\frac{\ddot{a}(t)}{a(t)}, Q^{'}(t)=\frac{\ddot{A}(t)}{A(t)}.
\end{eqnarray}
\endnumparts

These entries are matched with the corresponding entries of Energy-Momentum tensor

\begin{eqnarray}\label{T}
\nonumber && \left[T^{\mu}_{\hspace{1mm}\nu}\right]=diag\left(\rho_{tot}c^{2},\underbrace{\mathcal{P},\mathcal{P},...},\underbrace{\mathcal{P^{'}},\mathcal{P^{'}},...}\right) \\[-7pt]
          && \hspace{38mm} {\scriptstyle n\,times}  \hspace{6mm} {\scriptstyle N\,times}
\end{eqnarray}

through the Einstein field equation

\begin{equation}\label{ET}
^{(n+N+1)}E^{\mu}_{\hspace{1mm}\nu}=\kappa_{n+N+1}\,^{(n+N+1)}T^{\mu}_{\hspace{1mm}\nu};
\end{equation}

where $\kappa_{n+N+1}=8\pi G_{n+N+1}/c^{4}$ is the gravitational coupling constant of the $n+N+1$-dimensional space-time, $\rho_{tot}=\rho+\rho^{'}$, and the primed letters indicate compact extra dimensions. $\rho^{'}$ is defined as the excess added density measured by the observer of the n+1-dimensional universe originated in the existence of the compact extra dimensions. Apearance of this added density is due to sharing a time-like coordinate held in common by the large and compact dimensions in the metric. It should be noted that while the metric is not warped, the pressure $\mathcal{P}$ carries no effect of the extra dimensions in itself, just the same as the independency between ${\mathcal P}_{1},{\mathcal P}_{2},{\mathcal P}_{3}$ in our FRW universe (here all equal to $\mathcal{P}$) which no one leaves an effect in the other. However, the field equations produce the Friedmann-like equations as follows

\numparts
\begin{eqnarray}\label{Feqs}
\nonumber && \frac{n(n-1)}{2}\frac{\dot{a}^{2}(t)+kc^{2}}{a^{2}(t)}+\frac{N(N-1)}{2}\frac{\dot{A}^{2}(t)+jc^{2}}{A^{2}(t)}\\
&& +nN\frac{\dot{a}(t)}{a(t)}\frac{\dot{A}(t)}{A(t)}=8\pi G_{D+1}\rho_{tot}, \label{Feqs first} \\
&& \frac{\ddot{a}(t)}{a(t)}\,+\,\frac{\ddot{A}(t)}{A(t)}=-\frac{8\pi G_{D+1}}{D-1}\left((D-2)\rho_{tot}+\frac{n\mathcal{P}+N\mathcal{P^{'}}}{c^{2}}\right), \label{Feqs second}
\end{eqnarray}
\endnumparts

where $D=n+N$.

\section{\label{sec3}Asymptotic Behavior of Friedmann-like Equations}

This section studies the behavior of field equations in two special case where our FRW universe is spatially flat, and(or) compact extra dimensions are static.

\subsection{\label{sec3a}Special Case: $k=j=0$}

This case uncouples (\ref{Feqs first}) and (\ref{Feqs second}) in to two independent second-order equations for the two scale factors as

\numparts
\begin{eqnarray}\label{Feq00}
\nonumber && (n-1)(D-1)c^{4}\left(\frac{\dot{a}}{a}\right)^{4}\\
\nonumber && +2(D-1)^{2}c^{2}\left((D-1)c^{2}\frac{\ddot{a}}{a}+8\pi G_{D+1}\mathcal{S}_{2N}\right)\left(\frac{\dot{a}}{a}\right)^{2}\\
&& -(N-1)\left((D-1)c^{2}\frac{\ddot{a}}{a}+8\pi G_{D+1}\mathcal{S}_{1N}\right)^{2}=0, \label{Feq00 first} \\
\nonumber && (N-1)(D-1)c^{4}\left(\frac{\dot{A}}{A}\right)^{4}\\
\nonumber && +2(D-1)^{2}c^{2}\left((D-1)c^{2}\frac{\ddot{A}}{A}+8\pi G_{D+1}\mathcal{S}^{'}_{2n}\right)\left(\frac{\dot{A}}{A}\right)^{2}\\
          && -(n-1)\left((D-1)c^{2}\frac{\ddot{A}}{A}+8\pi G_{D+1}\mathcal{S}^{'}_{1n}\right)^{2}=0, \label{Feq00 second}
\end{eqnarray}
\endnumparts

 where

\numparts
\begin{eqnarray}\label{S12}
&& \mathcal{S}_{1N}=N\mathcal{P}^{'}-(N-1)\mathcal{P}-\rho_{tot} c^{2},\\
&& \mathcal{S}_{2N}=N\mathcal{P}^{'}-(N-1)(\mathcal{P}-\rho_{tot} c^{2}),\\
&& \mathcal{S}^{'}_{1n}=n\mathcal{P}-(n-1)\mathcal{P}^{'}-\rho_{tot} c^{2},\\
&& \mathcal{S}^{'}_{2n}=n\mathcal{P}-(n-1)(\mathcal{P}^{'}-\rho_{tot} c^{2}).
\end{eqnarray}
\endnumparts

 (\ref{Feq00 first}) shows that even in the n+N+1-dimensional space-time, the acceleration may still behaves as a gravitational source as expected of principle of equivalence in general relativity.\\
In the current special case, (\ref{Feqs first}) may be rewritten as

\begin{equation}\label{Uniqe}
H^{'}=-\frac{nNH\pm\sqrt{\Delta}}{N(N-1)},
\end{equation}

where

\begin{equation}\label{Delta}
\Delta=N\left[n(D-1)H^{2}+16\pi G_{D+1}(N-1)\rho_{tot} \right],
\end{equation}

Thus, the uniqueness condition for $H^{'}$ implies $N=0$, which means no extra dimensions should exist, or

\begin{equation}\label{1a00}
\left(\frac{\dot{a}}{a}\right)^{2}=-\frac{16\pi G_{D+1}(N-1)}{n(D-1)}\rho_{tot}.
\end{equation}

This result is consistent with relation (3) of \cite{Mansouri} (by setting $N=0 , n=D$). Also, as expected, (\ref{1a00}) gives the first ordinary Friedmann equation for $N=0 , n=3$.\\
Inserting this equation in to (\ref{Feqs second}), yields the acceleration equation

\begin{equation}\label{2a00}
\frac{\ddot{a}}{a}=\frac{8\pi G_{D+1}}{n(D-1)c^{2}}\mathcal{S}_{3},
\end{equation}

where

\begin{equation}\label{S3}
\mathcal{S}_{3}=n\left((N-1)\mathcal{P}-N\mathcal{P}^{'}\right)-\left(n+2(N-1)\right)\rho_{tot} c^{2},
\end{equation}

The equations for $A(t)$ may also get derived by applying the following transformations to the above energy and acceleration equations of $a(t)$:

\begin{equation}\label{S3}
a(t) \rightarrow A(t), \, N\rightarrow n, \, \mathcal{P}(t) \rightarrow \mathcal{P}^{'}(t).
\end{equation}

A remarkable note coming out of (\ref{1a00}) is that there is a distinguished case $N=1$ for the number of extra dimensions, in which our FRW universe has to be static in order to fulfill the field equations. Furthermore, if $N>1$, the expression $(N-1)\mathcal{P}-N\mathcal{P}^{'}$ in the source $\mathcal{S}_{3}$ appears as a pressure difference.

\subsection{\label{sec3b}Special Case: $\dot{A}=0$}

In this special case, field equations implies

\begin{equation}\label{ConstA}
A=jc^{2}\left[-\frac{8\pi G_{D+1}}{(N-1)(D-1)}\mathcal{S}^{'}_{1n}\right]^{-1/2},
\end{equation}

This equality contains two facts: \emph{Firstly}, if $j=0 \,or\, N=1$, existence of static extra dimensions is improbable unless an equilibrium holds as $\mathcal{S}^{'}_{1n}=0$ between the large scale dimensions and the extra ones; and \emph{Secondly}, if $j\neq0 \,and\, N\neq1$, static extra dimensions require $\mathcal{S}^{'}_{1n}=Const. \,$.

\section{\label{sec4}\emph{3+4+1}-Dimensional Space-time}

This section assumes dynamically compact extra dimensions to appear as a flat 4-dimensional part in the metric, equipped with a chart expressed by 4 coordinates as $(R,\psi,\vartheta,\varphi)$. This is an example in order to study the effects caused due to $N>3$. The metric is introduced as

\begin{eqnarray}\label{metric7+1}
&& ds^2=c^{2}dt^{2}-a^{2}(t)\left(\frac{dr^{2}}{1-kr^{2}}+r^{2}d\Omega^{2}\right)-A^{2}(t)\left(dR^{2}+R^{2}d\Omega^{'2}\right).
\end{eqnarray}

where

\numparts
\begin{eqnarray}\label{Omega}
&& d\Omega^{2}=d\theta^{2}+sin^{2}\theta \, d\phi^{2} \\
&& d\Omega^{'2}=d\psi^{2}+sin^{2}\psi \, d\vartheta^{2}+sin^{2}\psi \, sin^{2}\vartheta \, d\varphi^{2}.
\end{eqnarray}
\endnumparts

According to (\ref{Etensor first})-(\ref{Etensor third}), field equations of this case becomes

\numparts
\begin{eqnarray}\label{Feq7+1}
&& \frac{\dot{a}^{2}}{a^{2}}+k \frac{c^{2}}{a^{2}}+2\left(2\frac{\dot{a}}{a}\frac{\dot{A}}{A}+\frac{\dot{A}^{2}}{A^{2}}\right)=\frac{8\pi G_{8}}{3}\rho_{tot}, \label{Feq7+1 first} \\
&& \frac{\ddot{a}}{a}+2\left(\frac{\ddot{A}}{A}+\frac{\dot{a}}{a}\frac{\dot{A}}{A}+\frac{\dot{A}^{2}}{A^{2}}\right)=-\frac{4\pi G_{8}}{3}\left(\rho_{tot}+\frac{3\mathcal{P}}{c^{2}}\right), \label{Feq7+1 second} \\
&& \frac{\ddot{a}}{a}+\frac{\ddot{A}}{A}-\frac{\dot{a}}{a}\frac{\dot{A}}{A}-\frac{\dot{A}^{2}}{A^{2}}=-\frac{8\pi G_{8}}{3}\left(\rho_{tot}+\frac{\mathcal{P}^{'}}{c^{2}}\right) \label{Feq7+1 third}
\end{eqnarray}
\endnumparts

where $\rho_{tot}, \mathcal{P}, \mathcal{P}^{'}$ satisfy the following continuity equation:

\begin{equation}\label{Continuty7+1}
4\left(\rho_{tot}+\frac{\mathcal{P}^{'}}{c^{2}}\right)H^{'}+3\left(\rho_{tot}+\frac{\mathcal{P}}{c^{2}}\right)H+\dot{\rho}_{tot}=0.
\end{equation}

One can easily prove that this continuity equation together with the known fluid equation of the 3+1-dimensional FRW cosmos imply that $\rho^{'}\simeq\rho^{'}_{0}\,a^{-3}$, assuming $O(H^{'2}) \ll H$. This means, discussing a 3+4+1-dimensional space-time, if one requests continuity equation to be also held in our 3+1-dimensional FRW universe, the request may be fulfilled just when the excess density also behaves like the baryonic matter with respect to the scale factor of large-scale dimensions. Clearly, this result is consistent with what one expects of the notion of dark matter. Another interesting result coming out of (\ref{Feq7+1 first})-(\ref{Feq7+1 third}) is a combined equation of state as

\begin{equation}\label{State7+1}
\left(3G-2G_{8}\right)\rho c^{2}-2G_{8}\rho^{'}c^{2}=3G_{8}\left(2\mathcal{P}^{'}-\mathcal{P}\right).
\end{equation}

By considering equation of states $\mathcal{P}=\omega \rho^{'}c^{2}, \,\, \mathcal{P}^{'}=\omega^{'} \rho^{'}c^{2}$ one have

\begin{equation}\label{rhoG}
\frac{\rho}{\rho^{'}}=\frac{2G_{8}(3\omega^{'}+1)}{3G+(3\omega-2)G_{8}},
\end{equation}

which shows that the ratio of densities and the thermodynamical behavior of gravitational sources related to large and compact dimensions together, determine the ratio of their structural constants. \\
The following parts of this section studies the 3+4+1-dimensional space-time in some other special cases.

\subsection{\label{sec4a}Special Case: $\ddot{A}=0 , \, O(H^{'2})\ll H$}

Neglecting $\dot{A}^{2}/A^{2}$ in comparison with $\dot{A}/A$ and $\dot{a}/a$, and by eliminating $\ddot{a}/a$ between (\ref{Feq7+1 second}) and (\ref{Feq7+1 third}), one earns $(\dot{a}/a)(\dot{A}/A)$ in terms of $\rho_{tot},\mathcal{P}$ and $\mathcal{P}^{'}$. Substituting this expression for $(\dot{a}/a)(\dot{A}/A)$ into (\ref{Feq7+1 first}), one gets:

\begin{equation}\label{Special1}
\frac{\dot{a}^{2}}{a^{2}}+k\frac{c^{2}}{a^{2}}=\frac{8\pi G_{8}}{9}\left(2\rho_{tot}+\frac{3\mathcal{P}-2\mathcal{P}^{'}}{c^{2}}\right),
\end{equation}

which obviously shows how the pressure difference can act as an excess density from the perspective of a large-scale observer in the FRW universe.

\subsection{\label{sec4b}Flat Empty Universe: $\ddot{A}=0 , \, k=j=0 , \, \rho=\rho{'}=0$}

In this case, (\ref{Feq7+1 first}) and (\ref{Feq7+1 second}) reduce to the following linear equations

\begin{equation}\label{Special2}
\frac{\dot{a}}{a}+(2+\sqrt{2})\frac{\dot{A}}{A}=0, \hspace{3mm} \frac{\ddot{a}}{a}+2\frac{\dot{a}}{a}\frac{\dot{A}}{A}=0,
\end{equation}

Omitting $\frac{\dot{A}}{A}$ between the two, results in a non-linear differential equation as $\dot{a}^{2}=(1\pm\frac{\sqrt{2}}{2})a\ddot{a}$ with a pair of solutions in the form $a_{\pm}(t)=(C_{1}t+C_{2})^{1\pm\sqrt{2}}$. According to the conditions $a(0)=0, a(t_{0})=1$, constants of integration $C_{1}, C_{2}$ are equal to $t_{0}^{-1}, 0$, respectively; $t_{0}$ being the age of cosmos in the model. A remarkable note here is about the Hubble parameter which has the same form for both of solutions as $const.t_{0}/t$, with the same behavior as the Hubble parameter of a flat radiation-dominant FRW universe in a 3+1-dimensional space-time. But this is interesting that this behavior is reached for a totally empty universe which does not even contain any radiation. This issue shows that when the FRW observer studies the behavior of the Hubble parameter, the extra dimensions might appear as a radiation-like source of gravity. The two mentioned solutions coincide at $t=t_{0}$, and $a_{+}(t)$ represents an expanding FRW universe.

\subsection{\label{sec4c}Special case: $\ddot{A}=0 , \, k=j=0 , \, \rho{'}=0, \, \rho=\rho_{\Lambda}\equiv\rho_{\Lambda0}$}

This case yields a special solution as a de-Sitter space-time in the form $a(t)=a_{0}e^{\sqrt{3-\frac{16\pi G_{8}}{3}\rho_{\Lambda0}}\,t}$. Here, $\rho_{\Lambda0}$ is the density attributed to cosmological constant at the present time. The reality of the solution implies that $\frac{G_{8}}{G}\leq\frac{3}{2H_{0}^{2}\Omega_{\Lambda0}}$, $\Omega$ being the density parameter. Using the latest values reported by WMAP, the inequality means: $G_{8}\leq(\sim4.5\times10^{35})G$.

\subsection{\label{sec4d}Special Case: $k=j=0 , \, H^{'}=Const.\equiv H^{'}_{0}, \, \rho=\rho_{m}$}

Assume the 3+1-dimensional FRW universe only contains baryonic matter obeying the equation of state $\rho_{m}=\rho_{m0}a^{-3}$, where $\rho_{m0}=\rho_{m}(t_{0})$ and it is set $a(t_{0})=1$. Then, one can show that if the excess density behaves as $\rho^{'}=\rho^{'}_{0}(1+a^{-2})^{1/2}$, the information of the extra dimensions appears as a density of cosmological constant ($\rho_{\Lambda}$) in Friedmann energy equation; in a way that if $\rho^{'}_{0}=NnH^{'}_{0}\sqrt{\frac{\rho_{\Lambda}}{24\pi G}}$, then one has $\left(\frac{\dot{a}}{a}\right)^{2}=\frac{8\pi G}{3}(\rho_{m}+\rho_{\Lambda})$.

\section{\label{sec5}The $N=1$ Case}

A glance on (\ref{Etensor}) clearly shows that $N=1$ is a distinguished case from all other dimensionalities which the extra dimensions may have. This case reaches to interesting results such as follows.
Field equations for the discussing case are reached by setting $N=1$ in (\ref{Etensor first})-(\ref{Etensor third}), as

\numparts
\begin{eqnarray}\label{FeqsN=1}
&& H_{k}^{2}=\frac{2}{n(n-1)}\times\left[\frac{8\pi G_{D+1}}{n}\left((n-1)\rho_{tot}+\frac{n\mathcal{P}-(n-1)\mathcal{P}^{'}}{c^{2}}\right)+Q^{'}\right],\\
&& Q=\frac{1}{n}\times\left[-\frac{8\pi G_{D+1}}{n}\left((n-1)\rho_{tot}+\frac{n\mathcal{P}-\mathcal{P}^{'}}{c^{2}}\right)-Q^{'}\right].
\end{eqnarray}
\endnumparts

An amazing consequence of these equalities is that as $n=3$, expression $-\frac{16\pi G_{D+1}}{3c^{4}}\mathcal{P}^{'}+\frac{Q^{'}}{c^{2}}$ may play the role of the cosmological constant $\Lambda$ in Friedmann equations, if the following relation is satisfied for the acceleration of the extra dimensions:

\begin{equation}\label{Qprime}
Q^{'}=-\frac{4\pi G_{D+1}}{3}\left(\rho_{tot}-\frac{3\mathcal{P}^{'}}{c^{2}}\right).
\end{equation}

\section{\label{sec6}Conclusions}

The entries of the diagonal Einstein tensor have been obtained for a extra-dimensional anisotropic metric of the form (\ref{metric}). Field equations are investigated to reach a pair of coupled equations for the large scale factor $a(t)$ of our FRW universe and $A(t)$ of the compact extra part. \\
Uncoupling the equations in the case of spatially flat cosmos in section \ref{sec3a} has exhibited the form in which gravitational sources are added to the accelerations, respecting equivalence principle. Three interesting results were revealed in the special case of flat spatial parts: Firstly, if extra dimensions exist, implying $a(t)$ to fulfill the Friedmann equations is actually implying $H^{'}(t)$ ($=\frac{\dot{A}(t)}{A(t)}$) to be unique; Secondly, if only one extra dimension exists, then our FRW universe has to be static; and thirdly, the pressure difference between the large and compact dimensions behaves as an extra source of gravitation in the large-scale universe. \\
Investigation of a static compact part in section \ref{sec3b} has showed that a constraint as $\mathcal{S}^{'}_{1n}=Const. \,$ is required for our FRW cosmos in order to be static; which in a situation such as a flat or one-dimensional extra part, the $Const. \,$ is null. \\
In the next step, the 3+4+1-dimensional case has been studied in section \ref{sec4}, as an example with amazing results. It was found that the continuity equations necessitate the excess density to behave nearly as a baryonic matter in the manner of its equation of state; which is what one expects of the concept of dark matter. \\
Other conditions have also been searched within the 3+4+1-dimensional case in order to reach physical outcomes. Section \ref{sec4b} has showed that in an empty flat FRW universe, as a sub-space of a 3+4+1-dimensional manifold, the Hubble parameter behaves in the same way as a flat radiation-dominated FRW universe alone.This means the extra dimensions might have the potential to treat as a radiation-like gravitational source in some manners. Section \ref{sec4c} has discussed a flat $\Lambda$-dominated FRW universe when the extra dimensions apear as an empty flat 4-dimensional part with zero acceleration (Here, emptiness for extra dimensions means that they leave no effect such as $\rho'$ in the density). The discussion has calculated an upper limit for the higher-dimensional gravitational constant $G_{8}$ with respect to the Newtonian constant of gravity as $G_{8}\leq(\sim4.5\times10^{35})G$. Next Section, \ref{sec4d}, has briefly expressed the situation in which the compact extra part may act as a dark energy in the large-scale FRW universe. The same notion of dark energy has been studied again in section \ref{sec5} more generally, in the remarkable case which only one extra dimension exist. It results in a fact that the expression $-\frac{16\pi G_{D+1}}{3c^{4}}\mathcal{P}^{'}+\frac{Q^{'}}{c^{2}}$ may play the role of the cosmological constant $\Lambda$ in Friedmann equations of our FRW universe.





\section*{Acknowledgment}
This work was funded by the Research Institute for Astronomy and Astrophysics of Maragha (RIAAM) through a National Institute grant 1/4165-116.


\end{document}